# Problems and Prospects for Intimate Musical Control of Computers

David Wessel and Matthew Wright
Center for New Music and Audio Technologies
Department of Music
University of California Berkeley
Berkeley, California 94720
1 510 643 9990
{wessel, matt}@cnmat.berkeley.edu

### Abstract
In this paper we describe our efforts towards the development of live performance computer-based musical instrumentation. Our design criteria include initial ease of use coupled with a long term potential for virtuosity, minimal and low variance latency, and clear and simple strategies for programming the relationship between gesture and musical result. We present custom controllers and unique adaptations of standard gestural interfaces, a programmable connectivity processor, a communications protocol called Open Sound Control (OSC), and a variety of metaphors for musical control. We further describe applications of our technology to a variety of real musical performances and directions for future research.

### Introduction
When asked what musical instrument they play, there are not many computer music practitioners who would respond spontaneously with "I play the computer." Why not? In this report we examine the problems associated with the notion of the computer as musical instrument and the prospects for their solution.

Here at the onset it would be useful to consider some of the special features that computer technology brings to musical instrumentation. Most traditional acoustic instruments such as strings, woodwinds, brass, and percussion place the performer in direct contract with the physical sound production mechanism. Strings are plucked or bowed, tubes are blown, and surfaces are struck. Here the performer's gesture plays a direct role in exciting the acoustic mechanism. With the piano and organ the connection between gesture and sound is mediated by a mechanical linkage and in some modern organs by an electrical connection. But the relation between the gesture and the acoustic event remains pretty much in what one might call a *one gesture to one acoustic event* paradigm.

When sensors are used to capture gestures and a computing element is used to generate the sound, a staggering range of possibilities become available. Sadly but understandably, the electronic music instrument industry with its insistence on standard keyboard controllers maintains the traditional paradigm. Musical instruments and their gestural interfaces make their way into common use or not for a variety of reasons most of which are social in character. These more sociological aspects like the development or not of a repertoire for the instrument are beyond the scope of this paper. Here we will concentrate on factors such as ease of use, potential for development of skill, reactive behavior, and coherence of the cognitive model for control.

In the figure below we provide a conceptual framework for our controller research and development. Our human performer has intentions to produce a certain musical result. These intentions are communicated to the body's sensorimotor system ("motor program"). Parameters are sensed from the body at the gestural interface. These parameters are then passed to controller software that conditions, tracks, and maps them to the algorithms that generate the musical material. Admittedly this diagram is schematic and incomplete. One aspect that is not well captured by it is the way in which performers' intentions are elaborated upon by discovery of new possibilities afforded by the instrument. Experimental and otherwise exploratory intentions are certainly dear to the authors. We find that this albeit schematic framework allows us to view the roles of human motor learning, controller mapping, and generative software as an overall adaptive system [1].

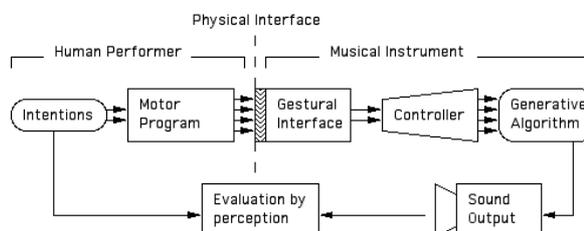

Unlike the *one gesture to one acoustic event* paradigm our framework allows for generative algorithms to produce complex musical structures consisting of many events. One of our central metaphors for musical control is that of *driving* or *flying* about in a space of musical processes. Gestures move through time as do the musical processes.

### Low entry fee with no ceiling on virtuosity
Getting started with a computer-based instrument should be relatively easy but this early stage ease-of-use should not stand in the way of the continued development of musical expressivity. Most of the traditional acoustical musical instruments are not easy to play at first but do afford the development of a high degree of musicality. On the other hand many of the simple-to-use computer interfaces proposed for musical control seem, after even a brief period of use, to have a toy-like character and do not invite continued musical evolution.

Is the low entry fee with no ceiling on virtuosity an impossible dream? We think not and argue that a high degree of control intimacy can be attained with compelling control metaphors, reactive low latency variance systems, and proper treatment of gestures that are continuous functions of time. With the potential for control intimacy assured by the instrument, musicians will be much more inclined to the continued development of performance skill and personal style.

**Latency requirements for control intimacy**

Few practitioners of live performance computer music would deny that low latency is essential. Just how low is the subject of considerable debate. We place the acceptable upper bound on the computer's audible reaction to gesture at 10 milliseconds (ms) and the systems described in this paper provide for measured [2] latencies nearer 7 ms.

Low variation of latency is critical and we argue that the range of variation should not exceed 1 ms. Grace-note-generated timbres as in flams can be controlled by percussionists with temporal precision of less than 1 ms. This is accomplished by controlling the relative distance of the sticks from head during the stroke. Timbral changes in the flams begin to become audible when the variations in the time between the grace note and the primary note exceed 1 ms. Psychoacoustic experiments on temporal auditory acuity provide striking evidence for this criterion [3, 4]. As we will argue below, prospects for the solution to the latency variation problem can be resolved by using time tags or by treating gestures as continuous signals tightly synchronized with the audio I/O stream.

**Discrete Event Versus Continuous Control**

MIDI is a discrete event protocol. MIDI events turn notes on and off and update changes in controller values. MIDI events are almost never synchronized with digital audio samples. Furthermore, MIDI provides no mechanism for atomic updates. Chords are always arpeggios and even when MIDI events are time tagged at the input of a synthesizer they arrive as a sequence. Moore [5], McMillen [6], and Wright [7] provide numerous examples of the dysfunction of MIDI. Much of this dysfunction is addressed by the Open Sound Control (OSC) protocol described below.

Many musical gestures are continuous functions of time and should be treated as such, for example, the position along the string of a finger on a violinist's left hand. The new generation of software synthesis systems such as Max/MSP www.cycling74.com, SuperCollider www.audiosynth.com, PD http://www-crca.ucsd.edu/~msp/software.html, and Open Sound World (OSW) www.cnmat.Berkeley.EDU/OSW provide for multi-rate signal processing. In these programming environments it is quite natural to treat gestures with a sample-synchronous signal processing approach. CNMAT's connectivity processor [8] described below provides a mechanism for getting continuous gestures into the computer in a manner that is very tightly synchronized with the audio sample stream. With this system we demonstrate a significant increase in control intimacy.

**Open Sound Control (OSC)**

Open Sound Control is a discrete event protocol for communication among controllers, computers, sound synthesizers, and other multimedia devices that is optimized for modern networking technology. Entities within a system are addressed individually by an open-ended URL-style symbolic naming scheme that includes a powerful pattern matching language to specify multiple recipients of a single message. We provide high-resolution time tags and a mechanism for specifying groups of messages whose effects are to occur simultaneously. Time tags allow one to implement a scheduling discipline [9] that reduces jitter by trading it for latency.

OSC's use of symbolic names simplifies controller mapping and its hierarchical name space helps in the management of complexity. There is also a mechanism for dynamically querying an OSC system to find out its capabilities and documentation of its features.

OSC has been integrated into Max/MSP (by Matthew Wright), Csound www.csounds.com (by Stefan Kersten and Nicola Bernardini) SuperCollider (by James McCartney), and OSW (by Amar Chaudhury). It has been used in a variety of contexts involving controllers. See www.cnmat.berkeley.edu/OSC for more detail and downloadable OSC software.

**A Programmable Connectivity Processor**

The conventional approach for communicating gesture and sound to real-time performance systems is to combine a microcontroller or DSP chip with A/D, D/A convertors and a network interface such as a MIDI serial controller. We have developed an alternative, more flexible approach that supports scalable implementations from a few channels of audio and gestures to hundreds of channels.

Our new system to address computer music and audio connectivity problems is based on integrating all digital functions on a single field programmable gate array (FPGA). All functions are determined by compiling high-level hardware descriptions (in VHDL) into FPGA configurations. This approach allows the considerable investment in developing the interface logic to each peripheral to be easily leveraged on a wide variety of FPGA's from different vendors and of different sizes. Since FPGA's are now available in sizes greater than a million gates, entire DSP and microcontrollers can also be integrated if required.

We have developed and tested VHDL descriptions for processing serial audio data for the SSI, S/PDIF, AES/EBU, AES-3, and ADAT industry standards. For gestures that are continuous we sample at submultiples of the audio rate and have VHDL modules for multichannel 8-bit, 12-bit and 16-bit A/D converters. We also provide modules for multiple MIDI input and output streams. Although such descriptions have been developed for proprietary systems, this library of modules represents the first complete, independent suite available in VHDL.

This suite makes possible some unusual cross codings such as embedding gestural data in audio streams, increasing temporal precision by exploiting isochronous data paths in the control processor. We sample continuous gestural signals at a submultiple of the audio sampling rate, multiplex the channels, and represent them as audio input signals. This allows us to get gestural signals into our software with the same low latencies as audio input, and guarantees that gestural and audio input signals will be tightly synchronized.

A novel module of particular importance in portable computer-music performance systems implements fast Ethernet from the hardware layer up through IP to the UDP protocol of TCP/IP. Because of the importance of Internet performance, Fast Ethernet implementations are extremely reliable and finely tuned on all modern operating systems.

A key feature of the connectivity processor is the analog subsystem for continuous gesture acquisition. We currently provide for 32 channels of analog-to-digital conversion. Voltage ranges of the converters are selectable as are the sampling rates which are constrained to be integer divisions of the audio rate and the appropriate antialiasing filter cutoff frequencies.

Our system combines VHDL connectivity modules that multiplexes 8 channels of bidirectional audio, MIDI, S/PDIF and transduced gestures into UDP packets which are exchanged with a portable computer using new, customized ASIO drivers in Max/MSP. See www.cnmat.berkeley.edu/ICMC2000 for a more detailed description of CNMAT's connectivity processor.

**Musical Control Structures for Standard Gestural Controllers**
Throughout history, people have adapted whatever objects were in their environment into musical instruments. The computer industry has invested significant resources in creating broadly available, low cost gestural controllers without any musical application in mind; thoughtful adaptation of these controllers for music is a fruitful yet overlooked route.

We find the latest incarnations of the venerable digitizing tablet (a.k.a. "artist's tablet") very interesting for musical control. Tablets offer accurate and fast absolute position sensing of cordless devices in three dimensions. Additionally, pressure, orientation, tilt and rotation estimates are available. The tablet we use allows for simultaneous sensing of two devices, usually one in each hand. This rich, multidimensional control information can be mapped to musical parameters in a variety of interesting ways.

The most direct kind of mapping associates a single synthesis parameter with each control dimension, for example, vertical position controlling loudness, horizontal position controlling pitch, etc. This kind of mapping proved to be musically unsatisfying, exhibiting the toy-like characteristic that does not allow for the development of virtuosity.

More interesting interfaces define regions of the tablet associated with particular behaviors. For example, one region might consist of a grid providing access to a large palette of musical material, while other regions represent musical processes that can operate on selected musical material. Repeating rhythmic cycles can be represented graphically on a region of the tablet, and sonic events can be placed at particular time points within the cycle [10].

We have created software in the Max/MSP environment that we use to develop control structures for the two-handed digitizing tablet. Examples include navigation in timbre space, multidimensional synthesis control, note stream synthesis, and emulations of the gestures of strumming, plucking and bowing strings. We also developed an interactive musical installation that uses two joystick controllers.

**Some Custom Controllers**
At CNMAT we have developed applications for variety of custom controllers, these include Don Buchla's *Thunder* and *Lightning* www.buchla.com, Tactex controllers www.tactex.com, Force Sensing Resistor (FSR) technology [11], and Piezo electric sensors in conjunction with percussion [12]. We currently have research projects underway that exploit a key feature of the previously described connectivity processor – namely the synchronization of control signals with the audio I/O stream. These projects include an organ keyboard with continuous sensing of each key position [13], a variety of micro-accelerometer projects, and new FSR devices.

In addition we have developed new sensor systems for multidimensional string motion [14] and have made some advances in extracting control signals from vocal sounds.

**Metaphors for Musical Control**
As suggested in the introduction, metaphors for control are central to our research agenda. We have found the work of George Lakoff and his collaborators [15, 16] on embodied cognition to be particularly applicable. They argue that abstract concepts like time and space and even the loftier concepts of mathematics are grounded in sensorimotor experience. We now present some of the metaphors that have inspired the development of our controller software.

**Drag and Drop**
The drag and drop metaphor is well known to users of the Apple Macintosh. An object is selected picked up and dropped upon a process. This is a natural application of Lakoff's movement and container metaphors. Our drag and drop system has been extended to the problem of the control of musical processes using the pen and tablet interface. Musical material is selected and then dropped onto a musical process.

One of the most critical features of any musical control system is a silencer, a mechanism that allows the performer to gracefully shut down a musical process. To this end we have the performer place the pen on the process and using a circular motion like the traditional copy editor's cursive delete sign the process is silenced at a rhythmically appropriate point in time. Other interfaces use the "eraser" end of the pen to silence processes.

### Scrubbing and its Variants

Sinusoidal models allow arbitrary time-scale manipulation without any change in pitch or spectral shape. We have built "scrubbing" interfaces for the tablet in which one dimension of the pen's position on the tablet maps to the time index of a sinusoidal model. Moving the pen gradually from left to right at the appropriate rate results in a resynthesis with the original temporal behavior, but any other gesture results in an alteration of the original. This interface allows a performer to play more arbitrary musical material, while preserving the fine continuous structure of the original input sounds. This kind of interface has been used in live performance contexts with classical Indo-Pakistani singing [17], trombone samples with expressive glissandi, and saxophone quartet material.

Other dimensions of the tablet sensing data, e.g., pressure, tilt, and vertical position, can be mapped to synthesis parameters such as loudness and spectral shape.

### Dipping

In the "dipping" metaphor the computer constantly generates musical material via a musical process, but this material is silent by default. The performer controls the volume of each process, e.g., using a poly-point touch-sensitive interface with the pressure in each region mapped to the volume of a corresponding process. Other gestural parameters control other parameters of the musical processes.

An advantage of this metaphor is that each musical event can be precisely timed, regardless of the latency or jitter of the gestural interface. Once a given process is made audible, its rhythm is not dependent on the performer in an event-by-event way.

This kind of interface is most satisfying when there are multiple simultaneous musical processes of different timbres, allowing the performer to orchestrate the result in real-time by selecting which processes will be heard.

### Acknowledgments

Special thanks to Rimas Avizienis, Adrian Freed, and Takahiko Suzuki for their work on the connectivity processor and to Gibson Guitar, DIMI, and the France Berkeley Fund for their generous support.